\documentclass[article,twocolumn,preprintnumbers,nofootinbib]{revtex4-1}
\usepackage{enumerate}
\usepackage{mathrsfs,amsmath,amssymb}
\usepackage{natbib}
\usepackage{graphicx}
\usepackage{cancel}
\usepackage[normalem]{ulem}
\usepackage[pdftex,bookmarks,linktocpage,pdfpagelabels,plainpages=false,hyperfigures,linkcolor=blue,citecolor=blue]{hyperref} 
\hypersetup{colorlinks=true}
\usepackage{stackrel}
\usepackage{pgfplots}
\usetikzlibrary{positioning}
\usetikzlibrary{snakes}
\usepackage{tikz-feynman}

\newcommand\be{\begin{equation}}
\newcommand\ee{\end{equation}}
\newcommand\bea{\begin{eqnarray}}
\newcommand\eea{\end{eqnarray}}

\def \Pslash {P\!\!\!\!/}

\begin{document}
\bibliographystyle{apsrev4-1}


\title{A new Direct Detection Strategy for the Cosmic Neutrino Background }
\author{Wei Chao}
\email{chaowei@bnu.edu.cn}
\affiliation{Center for Advanced Quantum Studies, Department of Physics, Beijing Normal University, Beijing, 100875, China}
\author{Jing-jing Feng}
\email{fengjj@mail.bnu.edu.cn}
\affiliation{Center for Advanced Quantum Studies, Department of Physics, Beijing Normal University, Beijing, 100875, China}
\author{Mingjie Jin}
\email{jinmj@bnu.edu.cn}
\affiliation{Center for Advanced Quantum Studies, Department of Physics, Beijing Normal University, Beijing, 100875, China}
\author{Tong Li}
\email{litong@nankai.edu.cn}
\affiliation{School of Physics, Nankai University, Tianjin, 300071, China}

\begin{abstract}
The direct detection of cosmic neutrino background (CNB) has been a longstanding  challenge in particle physics, due to its low number density and tiny neutrino masses.  In this work, we consider the spectrum of the CNB boosted by cosmic rays via  the neutrino self-interaction,  and calculate the event rate of the boosted CNB-plasmon scattering in term of the dielectric response, which accounts for in-medium screening effect of a condensed matter target.  This  can be taken as the new direct detection strategy for  the CNB in complementary to the traditional one, which captures the CNB on a $\beta$-unstable nucleus.  Our result shows that one can either see the event  of the CNB  for the exposure of per kg$\cdot$year, or puts a strong constraint on the neutrino self-interaction.   We further explore the background induced by the sub-MeV dark matter and the boosted super-light dark matter. 

\end{abstract}

\maketitle


\section{Introduction}\label{1}

The standard cosmological model predicts the existence of the Cosmic Neutrino Background (CNB) , which, once discovered, will be a milestone in cosmology and neutrino physics.  In the $\Lambda{\rm CDM}$ model, the present day temperature of  the CNB is about $1.95~{\rm K}$ and its average number density is about $56~{\rm cm}^{-3}$ for each helicity degree of freedom.  As a result, the direct detection of  the CNB  is extremely difficult when further taking into account its weak interaction with the Standard Model (SM) particles.  A traditional strategy for the direct detection of CNB is to  capture the CNB on a $\beta$-unstable nucleus via the inverse beta decay~\cite{Blennow:2008fh,Nussinov:2021zrj,Alvey:2021xmq,PTOLEMY:2018jst,Faessler:2016tjf,Li:2010sn,Arteaga:2017zxg,Zhang:2015wua,Lopez:1998aq}, $n+ \nu_e \to e + p$, which is a threshold-less reaction. The relevant searches have been carried out in KATRIN~\cite{KATRIN:2021fgc,KATRIN:2019yun} and PTOLEMY~\cite{PTOLEMY:2018jst} experiments.  In those experiments, one needs to pick out the signal of the CNB  from the continuous $\beta$ decay background. However, the two electron energy spectrums are split  by the active neutrino mass, i.e. $\Delta E_e \sim 2 m_\nu^{} $.  The current best limit on the absolute mass scale of neutrinos comes from the KATRIN experiment,  which gives $ m_\beta < 0.8~{\rm eV}$ at the 90\% C.L.~\cite{KATRIN:2021fgc,KATRIN:2019yun}. Besides,  the cosmological structure formation provides strong constraint on the sum of neutrino masses, and the Planck collaboration gives $\sum m_\nu <0.12~{\rm eV}$ at the 95\% C.L.~\cite{Planck:2018vyg}. Given the tiny neutrino masses, it is technologically challenging to distinguish the signal induced by the CNB from that given by the continuous $\beta$-decay, which suggests that this strategy only works in non-standard cosmology with sterile neutrinos or large neutrino masses~\cite{Alvey:2021xmq}. 

It is well-known that the neutrino physics is a new physics beyond the SM, and there might be new neutrino interactions in addition to the electroweak interaction mediated by the $Z$ and $W$ bosons.  Furthermore, some typical neutrino self-interactions have been proposed to relieve the Hubble tension problem~\cite{Blinov:2019gcj,Ghosh:2019tab,Schoneberg:2021qvd}, or to avoid the constraint of the X-ray observations  on the sterile neutrino dark matter (DM)~\cite{DeGouvea:2019wpf,Kelly:2020aks}.  In this letter, we first point out that the CNB may be boosted by cosmic rays given a new  neutrino interaction.  Theoretically, the boosted CNB,  that contains large kinetic energy, is possibly to be detected via  coherent elastic neutrino-nucleus (or neutrino-electron) scattering in various DM direct detection experiments.  However, the Sun is an active neutrino sources in the solar system and  countless energetic solar neutrinos pass through the Earth at every second, which will cover up any signal induced by the boosted CNB.  Looking into the flux of cosmic neutrinos in the Universe,  one can find that the spectrum is not continuous and there is a gap between solar neutrinos and the CNB~\cite{Vitagliano:2019yzm} fluxes, which can be complemented by the flux of the boosted CNB.  As a result,  the signal induced by the  $\mathcal{O}$(eV) scale boosted CNB is free from other cosmic neutrino background.

It is indeed very difficult to detect neutrinos of this energy range in any neutrino experiment or in a traditional DM direct detection equipment, which is originally designed to look for heavy DM, and  typical signals of which appear as charge, light or heat, induced by elastic or inelastic scattering of DM on the target.  Recently, a lot of attentions have been paid to the direct detection of sub-GeV or sub-MeV DM whose kinetic energy is not the largest scale.  It catalyzed the development of new direct detection techniques using condensed matter materials, which benefit from both  the low energy threshold and the high target density compared to the atomic or molecular detectors.  Considering that neutrinos is actually a hot DM, we propose the new strategy for the direct detection of the boosted CNB  using condensed matter materials via the neutrino-electron scattering. The scattering rate is given in terms of the dielectric response of a material, which has been extensively studied in condensed matter theory and experiment. Our results show that one can either detect the event rate induced by the boosted CNB, or put a strong constraint on the neutrino self-interaction. It should be mentioned that the scattering induced by hypothetical light DM will be the background of the direct detection of the CNB. We present several backgrounds induced by boosted DM and sub-MeV fermion DM. 

In the following section, we first show the spectrum of the boosted  CNB induced by exotic neutrino interactions. Then we calculate the event rate of the boosted CNB as the dielectric response.  We present DM induced background in section IV, and conclude in section V.

\section{Boosted CNB spectrum}

\begin{figure}[t]
  \centering
  \includegraphics[width=0.49\textwidth]{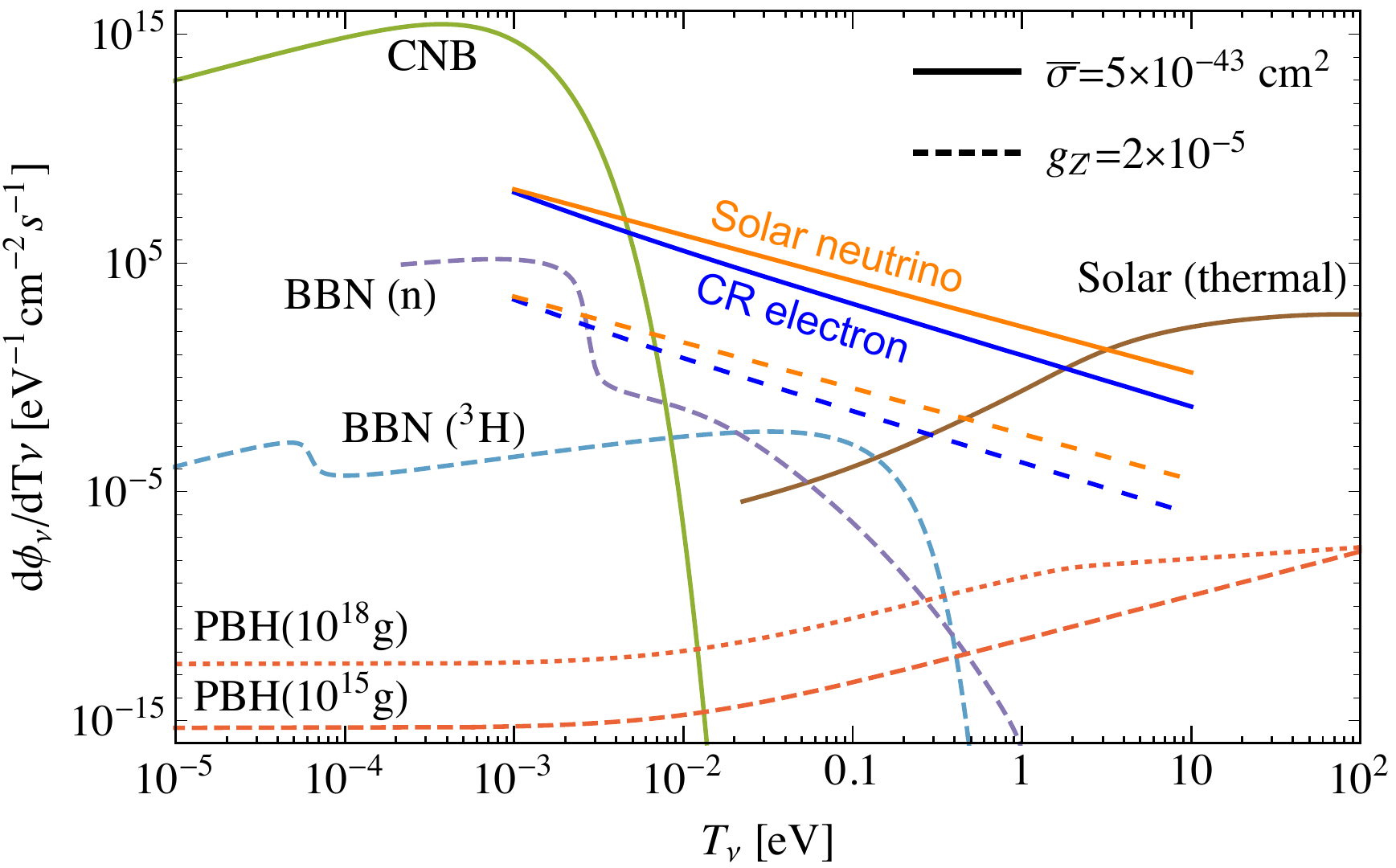}
\caption{The low energy neutrino flux as the function of the neutrino kinetic energy. The blue and orange lines represent the boosted CNB by electron and neutrino cosmic rays respectively, where the neutrino flux for a constant $\bar \sigma=5\times 10^{43} {\rm cm^2}$ (solid lines) and a constant $g_{Z'}=2\times10^{-5}$(dashed lines). Here we set the rest mass of CNB is $m_\nu=10^{-2}$ eV and assume $m_{Z'} \ll \alpha m_e$. The light green and brown solid lines denote the present-day CNB and thermal neutrino from the Sun. The light blue and purple dashed lines represent neutrino fluxes from $\beta$ decays at the BBN epoch. The red dashed and dotted lines indicate neutrinos from primordial black holes (PBHs)  evaporation with the mass $10^{15}$g and $10^{18}$g, respectively.}
\label{fig:nuflux}
\end{figure}
 
Active neutrinos decouple relativistically at about 2~${\rm MeV}$ in the early Universe. Its number density today is constrained by the CMB observable, namely the effective number of neutrino species.  The Planck collaboration gives  $N_{\rm eff}^{} =2.99\pm 0.17$ at the 68\% CL~\cite{Planck:2018vyg} and $\sum m_\nu <0.12 ~{\rm eV} $ at 95\% CL by using precise observation of CMB and Baryon Acoustic Oscillation data.  For direct detection of the CNB via its scattering off a target, it needs to be boosted by the energetic galactic cosmic rays, just like the case of boosted sub-GeV DM.  The differential flux of the boosted CNB  on the Earth can be written as 
 \bea
 {d \Phi_\nu (x)\over d T_\nu }=\int d^3 z d T_i {d \Phi_i(z)\over d T _i^{} }  \left. {d \bar \sigma\over d T_\nu^{} }\right|_{\theta=\theta_E^{} } {n_\nu (z) \over |z-x|^2} \label{nuflux}
 \eea
where $x$ is the position of the Earth and $z$ is the position where the scattering process takes place, $n_\nu^{}$ is the local number density of the CNB, $d\Phi_i/dT_i $ is the differential flux of the cosmic ray, $\bar \sigma_{\nu i}^{}$ is the scattering cross section of the CNB off the cosmic ray with reduced solid angle integration, $\theta_E$ is the scattering angle that keeps the boosted CNB towards the Earth. The minimal incoming energy required to obtain recoil energy $T_\nu$ for neutrino is $T^{\rm min}_i=(T_\nu/2 - m_i)(1\pm\sqrt{1+(2T_\nu(m_i+m_\nu)^2)/(m_\nu (2 m_i - T_\nu)^2)})$, where the sign $+(-)$ implies $T_\nu > 2 m_i (T_\nu < 2 m_i)$. If we further assume homogeneous CNB  and cosmic ray distributions, the differential flux can be simplified to the conventional one~\cite{Bringmann:2018cvk}, 
\bea
 {d \Phi_\nu \over d T_\nu } = \int \int\int {d \Omega d \ell d T_i \over 4 \pi}  { n_\nu } \cdot  { d \sigma_{ \nu i }^{} \over d T_\nu}  \cdot { d\Phi_i \over d T_i } 
  \eea
where  the full line-of-sight integration is performed out to $10$ kpc. We have neglected the neutrino flux attenuation due to propagation by assuming that the mean free path ( $d_\nu =1/n_\nu \sigma_\nu{} $) is much larger than $\ell$. 

In the SM, interactions between neutrinos and charged leptons (or neutrinos) are suppressed by heavy gauge bosons. To get a detectable spectrum of the  boosted CNB, we assume the existence of new neutrino interactions, which have been proposed to address various cosmological problems.   For a local U(1) gauge symmetry with the universal coupling $g_{Z^\prime}$ for charged lepton and neutrino interactions,  the elastic scattering cross section for the incident CR(electron~\cite{AMS:2014xys,Cummings:2016pdr,DAMPE:2017fbg,Cao:2020bwd} or solar neutrino~\cite{Vitagliano:2019yzm}) colliding with the CNB at rest  can be written as,
\begin{eqnarray}
\frac{d \sigma_{e\nu}}{d T_{\nu}}&=& \frac{g^4_{Z'}}{8\pi} \left\{ m_e^2(m_\nu-T_\nu) + m_e m_\nu (4 T_i - 2 T_\nu)  \right. \nonumber \\ 
&& \left.+ m_\nu T^2_\nu + m_\nu (T_i - T_\nu)^2 \right\} \nonumber \\ & & 
\times (2 m_e T_i + T^2_i)^{-1} (2 m_\nu T_\nu + m^2_{Z'})^{-2} \\
%
\frac{d \sigma_{\nu\nu}}{d T_{\nu}}&=& \frac{g^4_{Z'}}{2\pi} {m_\nu(m_\nu + T_i)^2(m_\nu T_i + m_{Z'}^2)^2  \over 2 m_\nu T_i + T^2_i }  \nonumber \\ 
&& (2 m_\nu T_\nu + m^2_{Z'})^{-2} (2 m_\nu(T_i - T_\nu) + m_{Z'}^2)^{-2}
\label{eq:diffsigma}
\end{eqnarray}
where we have assumed a universal neutrino mass $m_\nu$, $T_i= E_i - m_i$ being the kinetic energy of  the incoming particle and $T_\nu$ is the recoil energy of the CNB.  Alternatively, one can derive the spectrum of CNB for a reference cross section $\bar \sigma$, defined as $\bar \sigma \equiv g_{Z'}^4 m_\nu^2/(\pi (\alpha^2 m_e^2 + m_{Z'}^2)^2)$.

For  boosted CNB via neutrino self-interaction, sun-like stars provide main source of cosmic neutrino. One needs to sum up the entire stellar contributions in the Milk Way~\cite{Jho:2021rmn}
 \bea
{d\Phi_\nu^{\rm total} \over d T_\nu^{} }  = \int d^3 V n_{\rm star}^{} {d \Phi_\nu^{} \over d T_\nu}
\eea
where $n_{\rm star }$ is the distribution function of stellar in the galactic disk.

The Fig.~\ref{fig:nuflux} shows neutrino flux as the function of neutrino energy. The blue and orange lines represent the boosted CNB by electron and neutrino cosmic rays respectively, where the solid and dashed lines represent the neutrino flux for a constant $\bar \sigma=5\times 10^{-43} {\rm cm^2}$ (solid lines) and a constant $g_{Z'}=2\times10^{-5}$(dashed lines), respectively. Here we set the rest mass of CNB is $m_\nu=10^{-2}$ eV and assume $m_{Z'} \ll \alpha m_e$. The light green and brown solid lines denote the present-day CNB and thermal neutrino from the Sun. The light blue and purple dashed lines represent neutrino fluxes from $\beta$ decays at the Big Bang Nucleosynthesis (BBN) epoch. The red dashed and dotted lines indicate neutrinos from primordial black holes (PBHs)  evaporation with the mass $10^{15}$g and $10^{18}$g, respectively, in which we have assumed the following  fraction of the PBHs: $f_{\rm PBH}^{} =1$. One can conclude that the boosted CNB may cover the gap of neutrino flux around ${\cal O} (0.1)$ eV.  Thus any signal induced by neutrinos in this energy range will be a unique signature of the CNB.

\section{Event rate in condensed matter system}

\begin{figure}[t!]
  \centering
  \includegraphics[width=0.49\textwidth]{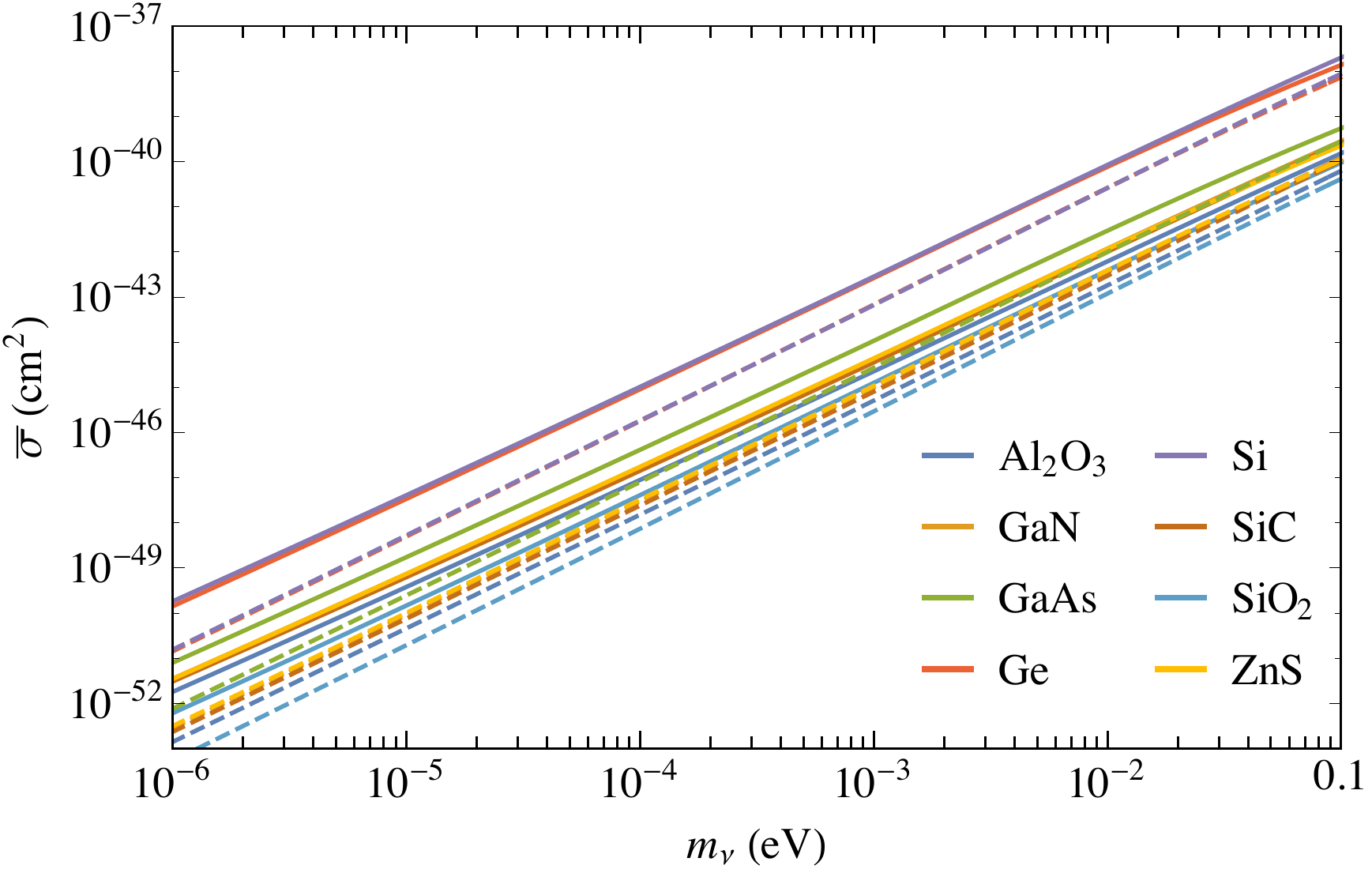}
\caption{The 95\% C.L. exclusion limit for cross section ($\bar \sigma$) in boosted CNB and electron scattering excited phonon signal with 1 kg-yr exposure. 
The colored lines indicate 8 solid state target materials: ${\rm Al_2O_3}$, ${\rm GaN}$, ${\rm GaAs}$, ${\rm Ge}$, ${\rm Si}$, ${\rm SiC}$, ${\rm SiO_2}$, ${\rm ZnS}$, where solid(dashed) type line indicates the constraint induced by the CR electron (solar neutrino). The ELF of each material is based on the Ref.~\cite{Knapen:2021bwg}.}
\label{fig:mzp_gzp}
\end{figure}

Recently, the condensed matter system as a new target for the direct detection of sub-GeV DM has been discussed explicitly.  The event rate induced by the DM-plasmon scattering are calculated using electronic wave functions with (time-dependent) density functional theory~\cite{Essig:2015cda, Trickle:2019nya, Griffin:2019mvc, Knapen:2021run, Hochberg:2021pkt, Griffin:2021znd, Knapen:2021bwg,Mitridate:2021ctr,Liang:2021zkg, Lasenby:2021wsc}. It is rather straightforward that when the deposited energy by the DM is larger than the  electron band gap, it can produce free electron excitations.  Alternatively, when the kinetic energy from incident particle is below the electron band gap of the target, the deposited energy can only produce collective phenomena in the target, i.e. plasmon, which can provide a much lower threshold for the detection.
Previous studies focus on the scattering between the non-relativistic DM and electron or nucleon. In this work, we provide the formula for the relativistic particle scattering off the target electron, and we will mainly consider the neutrino-plasmon scattering signal induced by the low energy boosted CNB. Our result may be applied to the  to the direct detection of the boosted DM.  

To calculate the event rate of the boosted CNB, we assume that the scattering  is mediated by a $Z'$ with universal gauge coupling $g_{Z'}$ to active neutrino and electron. Following the procedures in Refs.~\cite{Bellac:2011kqa, Altherr:1992mf},  the interaction rate for fermion can be written as,
\bea
\Gamma = {1\over 4 E} {\rm tr} \left[ (\Pslash+m_\nu ) \Sigma^{>} (P)\right] \label{xmaster}
\eea
where  $P=(E, p)$ being the momentum of fermion. $\Sigma^{>} (P)$ is the neutrino's in-medium cut self-energy given by the following Feynman diagram,
\bea
\begin{tikzpicture}
\draw[gray!60,fill=gray!60](2, -0.50) ellipse [x radius=0.5 cm, y radius=0.2cm];
\draw[-,ultra thick] (0,-1)--(4,-1);
\draw [-,snake=snake, ultra thick] (1,-1) -- (1.5, -0.5);
\draw [-,snake=snake, ultra thick] (2.5,-0.5) -- (3, -1);
\node[red, thick] at (0.5,-1.3) {$\nu$,~~$P$};
\node[red, thick] at (3.5,-1.3) {$\nu$,~~$P$};
\node[red, thick] at (1,-0.5) {$Z^\prime$};
\node[red, thick] at (2,0) {$e$};
\node[red, thick] at (2,-1.3) {$P-Q$};
\node[red, thick] at (3.2,-0.5) {$Z^\prime$,~~$Q$};
\end{tikzpicture} \nonumber
\eea
in which the ellipse represents the electron induced effect in the medium. A straight forward calculation gives
\bea
\Sigma^{>} (P) = g_Z^{\prime 2} \int {d^4 Q \over (2\pi)^4 } \gamma^\mu S_{0}^{>} (P-Q) \gamma^\nu D_{\mu\nu}^{>} (Q) 
\eea
where $S_0^{>} (P-Q)$ is the propagator of neutrino in vacuum, and $D_{\mu\nu}^{>}$ is the cut propagator of $Z^\prime$ in terms of free propagator plus the effect of the in-medium self-energy, that is $D_{\mu\nu}^{>} =D_{\mu\nu}^{>0} (Q) -2 (Q^2- M_{Z^\prime}^2)^{-2} {\rm sgn} (Q_0)[1 + f(Q_0)] {\rm Im} \left(\Pi_{\mu\nu}^{}  \right) $.  In our calculation, we neglect the $f(Q_0^{})$ term because the temperature of the medium is negligible. $\Pi_{\mu\nu}$ can be  further decomposed as transverse and longitudinal components, $\Pi_{\mu\nu}^T$ and $\Pi_{\mu\nu}^L$, which are related to the magnetic permeability and the electric permeability~\cite{Altherr:1992mf}.  Using the formula $Q^\mu_{}\Pi_{\mu\nu} (Q)=0$, implied by the current conservation, one can easily derive the expression of the interaction rate.

Convoluting the boosted CNB flux in Eq.~(\ref{nuflux}) with the transition rate in Eq.~(\ref{xmaster}) , the differential event rates per energy and per unit of exposure can be written as
\begin{eqnarray}
\frac{d R}{d \omega}&=& \frac{g^4_{Z'}}{(2\pi)^2\rho_T} \int d T_\nu \frac{d \Phi_\nu}{d T_\nu} \int \frac{dq }{E_\nu E_\nu'} \nonumber \\ 
&\times& {q^4 \over (q^2 + m_{Z^\prime}^2)^2}  \delta (E_\nu-E_\nu^\prime -\omega)\nonumber \\
&\times& \left(\frac{q^2}{2} + 2 E^2_\nu + 2m^2_\nu \right) {\rm Im} \left[ \frac{-1}{\varepsilon_L(q, \omega)} \right]
\label{eq:diffrate}
\end{eqnarray}
where $\rho_T$ is the target mass density, $E_\nu$ is the incident energy of the boosted CNB,  $E_\nu^\prime$ is the energy of the out-going neutrino, $\omega$ is the energy deposit in target, $\varepsilon_L(q,\omega)$ is the longitudinal dielectric function.   The last part of the Eq.~\ref{eq:diffrate} is also known as the energy loss function (ELF), which is the target dependent object and and can be calculated numerically in density functional theory. It should be mentioned that we have restricted ourselves to only consider non-magnetic materials when deriving the Eq.~(\ref{eq:diffrate}). A systematic study of neutrino(DM)-plasmon scattering rate in a magnetic material will be presented in a future work. 

For numerical analysis, we restrict to the following neutrino energy range $\mathcal{O}(0.01-0.1)$eV which merely covers the energy range of a target that can produce collective phenomena.  The advantage of searching for  the signal of the CNB in this energy range is that it is free from other neutrino backgrounds from celestial bodies, as can be seen from the Fig.~\ref{fig:nuflux}.  Alternatively, this choice only shows a conservative limit on the boosted CNB.  The Fig.~\ref{fig:mzp_gzp} shows the $(m_\nu-\bar \sigma)$ exclusion limit from the boosted CNB-plasmon scattering signal with 1 kg$\cdot$year exposure. 
The colored lines indicate the following eight  solid state target materials: ${\rm Al_2O_3}$, ${\rm GaN}$, ${\rm GaAs}$, ${\rm Ge}$, ${\rm Si}$, ${\rm SiC}$, ${\rm SiO_2}$, ${\rm ZnS}$, where solid(dashed) type line indicates the constraint induced by the CR electron (solar neutrino). We use the \textbf{DarkELF} package to calculate the plasmon excitations in the boosted CNB - electron scattering~\cite{Knapen:2021bwg}. The \textbf{DarkELF} package uses the ELF~\cite{Knapen:2021run, Hochberg:2021pkt} based on the first principles.

\section{Background and Discussion}

\begin{figure}[t]
  \centering
  \includegraphics[width=0.49\textwidth]{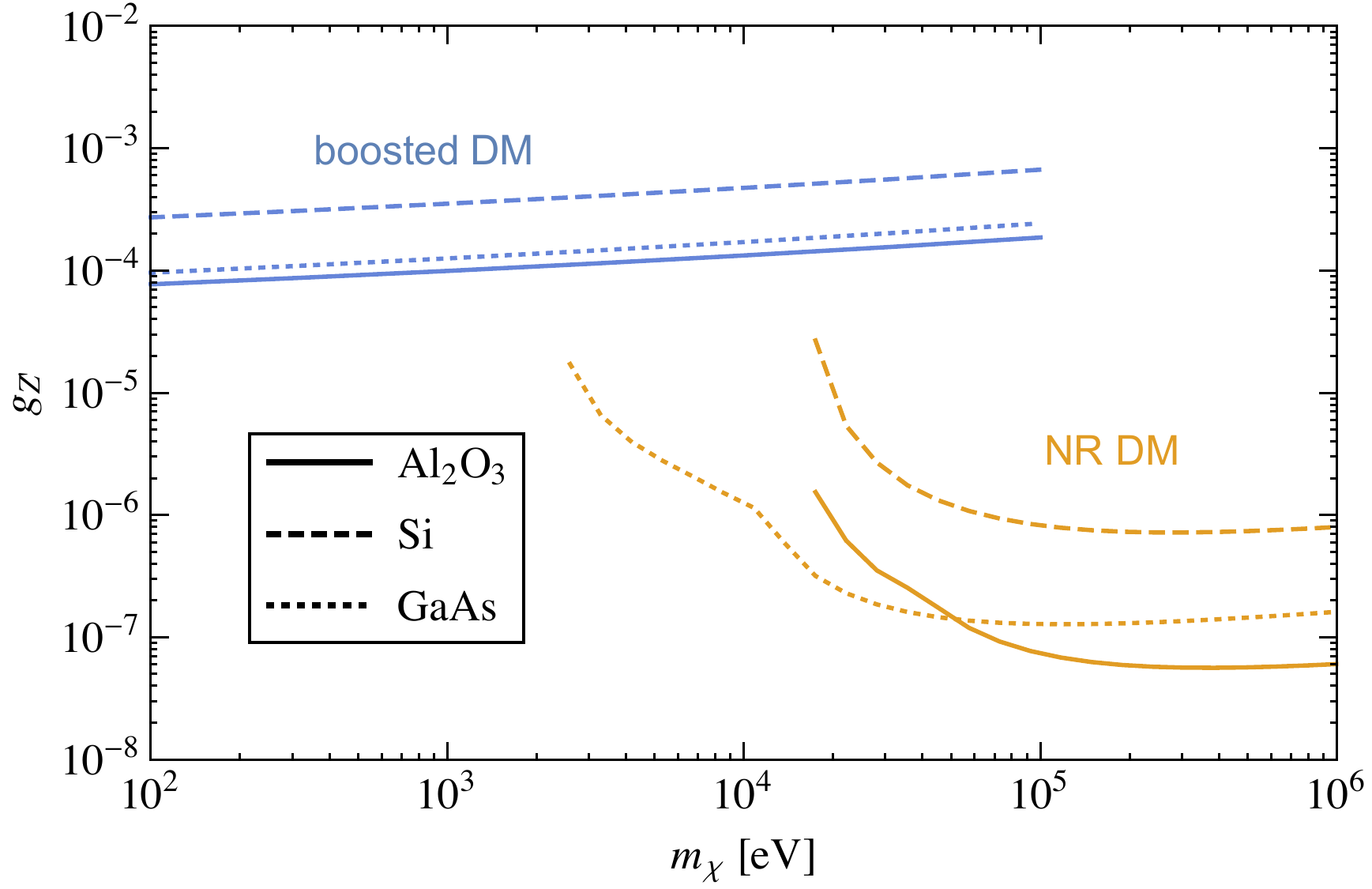} 
\caption{The exclusion limit for $g_{Z'}$ as a function of $m_\chi$. The solid, dashed and dotted lines denote constraints of the  target material ${\rm Al_2O_3}$, ${\rm Si}$ and ${\rm GaAs}$, respectively. The blue line describe the DM boosted by the CR electron. For comparison, the orange lines indicate the bounds in non-relativistic limits~\cite{Knapen:2021bwg}.}
\label{fig:dmlimit}
\end{figure}

As can be seen from the last section, a typical signal of the boosted CNB is plasmon/phonon excitations in a condensed matter system.  Actually, the same signal can be induced by  the hypothetical sub-MeV DM scattering off the electron on the target. In the past few years, there are vast studies on the calculation of DM event rate $dR_{DM}/d\omega$,  which relies on a reference cross section $\bar \sigma_e$, defined as $\bar \sigma_e=g_\chi^2 g_e^2 \mu_{\chi e}^2/(\pi (\alpha^2 m_e^2 + m_{\rm med}^2)^2)$.
In this section, we consider the signal of a light fermion DM by constraining the coupling constant of $g_{Z'}$,  as shown the Fig.~\ref{fig:dmlimit}. For boosted DM, we follow the procedure discussed in Ref.~\cite{Cao:2020bwd} to calculate the boosted DM flux and use Eq.~(\ref{eq:diffrate}) to calculate the event rate, where DM number density  $\rho_\chi/ m_\chi$ is used instead of the CNB number density $n_\nu$.


In Fig.~\ref{fig:dmlimit}, we choose three target materials: ${\rm Al_2O_3}$, ${\rm Si}$, ${\rm GaAs}$ and compare with the bounds in non-relativistic (NR) limit~\cite{Knapen:2021bwg}.
For the boosted DM case, the dominant energy of the DM is from its kinetic energy, which results in an approximately constant energy dependence in low mass range compared to NR DM, and moreover the bound declines as the mass decrease as the number density is large for light DM.

Back to constraint on the model presented in this paper, neutrino self-interaction is severely constrained by various decay processes~\cite{Kelly:2020pcy}, such as  Higgs invisible decay, $W/Z$ boson decays,  exotic meson decays, etc.  On the cosmological side, the studies of impacts of  neutrino exotic interaction on the BBN~\cite{Grohs:2020xxd,Blinov:2019gcj,Escudero:2019gzq} show that  there is a lower bound on the $Z^\prime$ mass  whenever it is thermalized in the early Universe.  Moreover, the new vector boson,  lighter than $100$~{MeV}, can be produced from neutrino scattering in the Supernova, which may modify the observed time scale of neutrino emission. Alternatively, cosmological constraints  more or less depend on the assumption of $\Lambda$CDM and the particle physics model, while our result in section III provides a new attempt of constraining these interactions in the terrestrial laboratory.

\section{Conclusion}

Direct detections of CNB has been a long standing challenge in the high energy physics. In this letter, we propose a new direct detection strategy of CNB by considering the boosted CNB spectrum induced by new neutrino interactions, which may cover the gap, $(0.01~{\rm eV},~1~{\rm eV})$ in the  cosmic neutrino flux spectrum.  The boosted CNB may scatter off the electron in condensed matter material and induces detectable collective phenomena. The non-observation of any signal within a given exposure will put constraint on the exotic neutrino interaction.  The same strategy is applicable  to the direct detection of super-light DM boosted by cosmic rays.  It should be mentioned that the model presented here is only a prototype, a further systematic study of neutrino interactions and their direct detection signals is needed.

\section*{Acknowledgments}
We thank to Profs. Zhi-zhong Xing, Shun Zhou, Yu-feng Li, Xun-jie Xu, Ka Shen, Li Wang, Jin-xing Zhang and Dr. Hai-jun Li for helpful discussions. This work was supported by the National Natural Science Foundation of China under grant No. 11775025 and No. 12175027. TL is supported by the National Natural Science Foundation of China (Grant No. 11975129, 12035008) and “the Fun- damental Research Funds for the Central Universities”, Nankai University (Grant No. 63196013).



\bibliography{CNB-BDM}


\end{document}